\documentclass[%
 aip,
 amsmath,amssymb,
]{revtex4-1}

\usepackage{graphicx}
\usepackage{dcolumn}
\usepackage{bm}

\usepackage[utf8]{inputenc}
\usepackage[T1]{fontenc}
\usepackage{mathptmx}
\usepackage{etoolbox}

\makeatletter
\def\@email#1#2{%
 \endgroup
 \patchcmd{\titleblock@produce}
  {\frontmatter@RRAPformat}
  {\frontmatter@RRAPformat{\produce@RRAP{*#1\href{mailto:#2}{#2}}}\frontmatter@RRAPformat}
  {}{}
}%
\makeatother
\begin{document}

\preprint{}

\title[]{Contact angle hysteresis can modulate the Newtonian rod climbing effect}
\author{Navin Kumar Chandra}

\author{Kaustuv Lahiri}

\author{Aloke Kumar}

\affiliation{ 
Department of Mechanical Engineering, Indian Institute of Science Bangalore, Karnataka, India, 560012
}%

\email{alokekumar@iisc.ac.in}

\date{\today}

\begin{abstract}
The present work investigates the role of contact angle hysteresis at the liquid-liquid-solid interface (LLS) on the rod climbing effect of two immiscible Newtonian liquids using experimental and numerical approaches. Experiments revealed that the final steady state contact angle, $\theta_{w}$ at the LLS interface varies with the rod rotation speed, $\omega$. For the present system, $\theta_{w}$ changes from $\sim$69$^{\circ}$ to $\sim$83$^{\circ}$ when the state of the rod is changed from static condition to rotating at 3.3 Hz. With further increase in $\omega$, the $\theta_{w}$ exceeds 90$^{\circ}$ which cannot be observed experimentally. It is inferred from the simulations that the input value of $\theta_{w}$ saturates and attains a constant value of $\sim$120$^{\circ}$ for $\omega>$ 5 Hz. Using numerical simulations, we demonstrate that this contact angle hysteresis must be considered for the correct prediction of the Newtonian rod climbing effect. Using the appropriate values of the contact angle in the boundary condition, an excellent quantitative match between the experiments and simulations is obtained in terms of- the climbing height, the threshold rod rotation speed for onset of climbing, and the shape of liquid-liquid interface. This resolves the discrepancy between the experiments and simulations in the existing literature where a constant value of the contact angle has been used for all speeds of rod rotation.
\end{abstract}
\maketitle

\section{\label{sec:level1}Introduction}

An interesting non-intuitive behaviour is displayed by the interface of two immiscible Newtonian liquids in the presence of a vertically immersed rotating rod. Beyond a certain rod rotation speed, the liquid-liquid interface shows a climbing in the vicinity of the rod. This non-intuitive phenomenon was first reported by \citet{bonn2004rod} and they named it as the Newtonian rod climbing effect. The essential criteria for climbing is that the lighter liquid should have higher viscosity otherwise a dip is observed instead of a climb \citep{bonn2004rod,zhao2017deformation}. The name "rod climbing effect" comes from a similar phenomenon classically observed at the air-liquid interface of viscoelastic liquids like polymer melts \citep{dealy1977weissenberg} and polymeric solutions \citep{weissenberg1947continuum,joseph1973free,chandra2021contact}. Although the stretched interface has a similar shape in both, the Newtonian and the viscoelastic rod climbing effect, the governing mechanism is completely different. In case of the viscoelastic liquids, rod climbing happens due to the dominance of the normal stress differences over the centrifugal force. However, in case of the Newtonian rod climbing effect, the inertial force due to secondary flows in the heavier liquid provides the necessary driving force for climbing of the interface \citep{bonn2004rod}. This mechanism of secondary flow-driven rising of liquid-liquid interface is confirmed by previous studies but in a different geometry. For instance, \citet{berman1978two} did early studies on the upheaval of the liquid-liquid interface in a spinning centrifuge tube. Similar phenomenon was reported by \citet{fujimoto2009topology} in a stationary cylindrical container with a circular rotating lid on top. The particular case of liquid-liquid interface deformation by a rotating rod is studied by \citet{zhao2017deformation} and \citet{bonn2004rod}. \citet{zhao2017deformation} demonstrated that rods of varied materials (Teflon, PVC, stainless steel) produced comparable magnitudes of climbing height. Based on this observation, the authors suggested that Newtonian rod climbing is mainly hydrodynamic in nature and does not depend on the wetting properties. They also supplemented their experiments with simulations which show a qualitative match, thereby establishing the fact that the Newtonian rod-climbing effect can be captured numerically through the Navier-Stokes equations using the Volume of Fluid (VOF) approach for multiphase formulation.

Existing literature does not account for the contact angle hysteresis (CAH) at the three-phase contact line (CL) of the liquid-liquid-solid interface in the rod climbing of Newtonian liquids. Although our previous work \citep{chandra2021contact} outlines the role of CAH due to CL pinning, but for an entirely different system of gas-liquid-solid interface in the rod climbing of viscoelastic liquid.
Dynamics of a liquid-liquid-solid CL as opposed to gas-liquid-solid CL, presents a more complex system because the interaction of the two liquids with the solid surface is comparable \citep{zheng2021effects}. In such systems CAH can be observed depending upon the roughness and wetting properties of the solid substrate \citep{fetzer2009dynamics,ramiasa2012nanoroughness,zanini2017universal}, surface energies of different materials \citep{hejazi2013contact}, and viscous dissipation due to flow near the three-phase CL \citep{fermigier1991experimental,seveno2011predicting}. In case of the presence of CAH, the contact angle for a given system can exhibit a range of steady state values. However, \citet{zhao2017deformation} modelled Newtonian rod climbing with a constant contact angle of 90$^{\circ}$ irrespective of the rod rotation speed, which is not true as we show in the present work. Although \citet{bonn2004rod} emphasized on the importance of the capillary force due to the interface curvature near the rod, they did not scrutinize the role of CAH on the Newtonian rod climbing. 
In the present work we provide experimental evidence for the existence of CAH in the Newtonian rod climbing effect shown by a stratified system of Silicone oil and DI water in the presence of a rotating stainless steel rod. The quantitative discrepancy between the numerical and experimental results of \citet{zhao2017deformation} can be reconciled by accounting for the change in the contact angle due to CAH. We show that, by using the appropriate value of contact angle prescribed as a boundary condition in the numerical simulations, different aspects of Newtonian rod climbing can be captured precisely. In particular, the climbing height, the threshold rod rotation speed for onset of climbing, and the shape of oil-water interface have been used as the quantifying parameters to test the proposed hypothesis.

\section{Materials and Methods}
\begin{figure}
\centering
\includegraphics[width=0.6\linewidth]{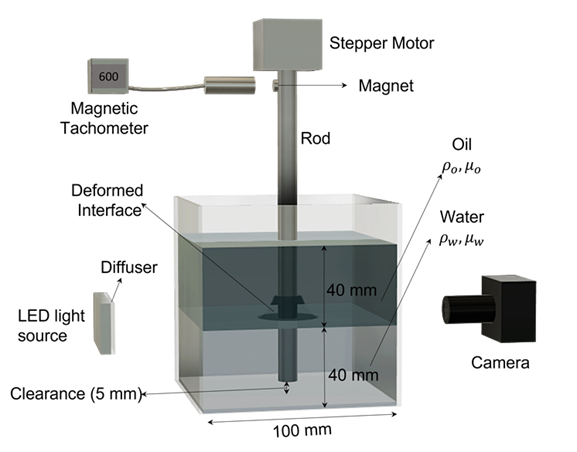}
\caption{Schematic diagram of the experimental setup. }
\label{fig:expsetup}
\end{figure}
The experimental setup consists of a vertical steel rod (diameter, $D \sim$ 10 mm) inserted into a cubical acrylic tank of edge length 100 mm, with transparent walls as shown in Figure \ref{fig:expsetup}. The rod rotation speed, $\omega$ is controlled precisely with a stepper motor connected at the upper end of the rod. $\omega$ is varied from 0 to 10 Hz in steps of 0.83 Hz. A magnetic tachometer continuously monitors $\omega$. It is observed that the fluctuations in $\omega$ from a constant set value is small ($\pm$ 0.05 Hz) compared to the range and the step increment of $\omega$ considered in the present study. Therefore, these fluctuations are neglected, and a constant value of rod rotation speed is reported. Despite a robust arrangement of mechanical bearings and couplings between the rod and the stepper motor, lateral vibrations of the rod is inevitable. However, the amplitude of lateral vibrations at the rod-water-oil interface location is very small ($\sim$150$\pm$25$\mu$m) compared to any other relevant length scale of the system. Moreover, the oil-water interface is symmetric in visual observation suggesting that the lateral vibration of the rod is insignificant for the purpose of present study.
To perform the experiments, the empty tank is first set in place with respect to the rod as shown in Figure \ref{fig:expsetup}. Then two immiscible Newtonian liquids are poured into the tank. The heavier liquid is DI water and the lighter one is Silicone oil. DI water is poured first after which Silicone oil is poured gently to prevent emulsification. 400 mL of liquid is poured to occupy a liquid column height of 40 mm for each liquid. A clearance of 5 mm is present between the tank bottom and the lower end of the rod. This geometry ensures that the length of rod dipped in each liquid is more than 25 mm, which is above the limit for the dip length to have any significant effect on the Newtonian rod climbing phenomenon\citep{zhao2017deformation}. The dynamic viscosity of DI water, $\mu_w \sim$ 1 mPa-s and silicone oil, $\mu_o \sim$ 355 mPa-s are measured using a cone-plate (40 mm diameter) geometry of a rheometer (Anton Paar, MCR 302). Interfacial tension at the oil-water interface, $\gamma \sim$ 34 mN/m is measured by using optical contact angle measuring and contour analysis systems (OCA 15EC, Dataphysics) by creating a pendant drop of water inside an oil bath. Densities of DI water, $\rho_w$ and Silicone oil, $\rho_o$ are not measured and their standard values are considered as 998 kg/m$^3$ and 968 kg/m$^3$ respectively. A DSLR camera (Model: D850, Nikon) and an LED light source are placed at two opposite faces of the liquid tank to track the shape of the oil-water interface near the rod. Images are captured using a macro lens (Sigma 105mm) in combination with the DSLR camera at high resolution ($\sim$25 $\mu$m/pixel) revealing some unique features of the Newtonian rod climbing which are not reported in the previous studies. Experiments are performed by slowly ramping $\omega$ from the static condition to attain a desired constant value. Sufficient time ($\sim$20 seconds) is allowed at constant $\omega$ for the oil-water interface to achieve a steady stable shape. Finally, the rod rotation is stopped abruptly, and the interfacial meniscus is allowed to recede and occupy the steady static shape. This whole cycle from initial static rod, to constant $\omega$ and the final static condition is considered as one dataset. A minimum of three datasets have been used for averaging and error analysis. Additionally, we employed particle image velocimetry (PIV) technique to measure the secondary flow velocity field in the meridional plane of water phase. PIV measurements are performed using glass spheres with mean particle size in the range $\sim$9-13 $\mu$m and density $\sim$ 1100 kg/m$^3$ (purchased from Sigma-Aldrich) as the tracer particle. Meridional plane of interest is illuminated by a laser sheet created with the help of a cylindrical lens. PIV recordings are done at 1000 frames per second using a high speed camera (Photron Fastcam mini AX100) in combination with a Navitar zoom lens (6.5X zoom 6000 series). PIV recordings are analyzed using PIVlab software available in MATLAB.

\section{Numerical Methods}
A finite volume based commercial computational fluid dynamics (CFD) code (ANSYS 2021-R2) is employed to perform the numerical simulations. The Volume of Fluid method (which is a surface tracking method) is applied to a fixed Eulerian mesh in order to solve the multiphase system. All the physical properties and variables are shared by the fields and their volume averaged values are employed at the interface. The interface is tracked by solving the transport equation for the aqueous phase volume fraction $\alpha_w$ varying between 0 and 1.

\begin{equation}
\frac{\partial{\alpha_{w}}}{\partial t} + \boldsymbol{u}.\boldsymbol{\nabla} \alpha_{w} = 0 
\label{eqn:eqn1}
\end{equation}
Where $\boldsymbol{u}$ stands for the velocity field and $t$ for time.
A single set of the Navier-Stokes equations was solved for both the fluids. 
\begin{equation}
\frac{\partial \rho}{\partial t} + \boldsymbol{\nabla}.\rho \boldsymbol{u} = 0
\label{eqn:eqn2}
\end{equation}
\begin{equation}
\frac{\partial \rho \boldsymbol{u}}{\partial t} + \boldsymbol{\nabla}.\rho \boldsymbol{uu} = \boldsymbol{-\nabla} p + \rho \boldsymbol{g} + \boldsymbol{\nabla}.\mu({\boldsymbol{\nabla u} + \boldsymbol{\nabla u ^ {T}}}) + \boldsymbol{{F_{\sigma}}}
\label{eqn:eqn3}
\end{equation}
The density is denoted by $\rho$, $p$ stands for the Pressure field, $\boldsymbol{g}$ stands for the gravitational acceleration vector, $\mu$ is the dynamic viscosity and $\boldsymbol{F_{\sigma}}$ stands for the surface tension force which is treated as a body force.  Equations \ref{eqn:eqn2} and \ref{eqn:eqn3} depend on the volume fractions through the characteristics of the density and viscosity. The volume weighted value of the properties are used in the computational cells having both of the phases.
\begin{equation}
\rho = \alpha_{w}\rho_{w} + (1 - \alpha_{w})\rho_{o}
\label{eqn:eqn4}
\end{equation}
\begin{equation}
\mu = \alpha_{w}\mu_{w} + (1 - \alpha_{w})\mu_{o}
\label{eqn:eqn5}
\end{equation}    
where the subscripts $o$ and $w$ stand for oil and water respectively.
The wall adhesion model is also employed which allows for defining the contact angle by taking the input parameter $\theta_w$ from the user. The corresponding equation for the normal vector, $\boldsymbol{n}$ of the interface adjacent to the wall is given by 
\begin{equation}
\boldsymbol{n} = \boldsymbol{n_{w}}cos(\theta_w) + \boldsymbol{n_{t}}sin(\theta_w) 
\end{equation}
where $\boldsymbol{n_{w}}$ is the normal vector to the wall, $\boldsymbol{n_{t}}$ is the tangential vector to the wall and $\theta_{w}$
is the wall adhesion angle. $\theta_{w}$ is an input based on experimental observations if $\theta_{w}<90^{o}$, and is obtained through an iterative process in case $\theta_{w}>90^{o}$. This provides a method for modulating the curvature, $\kappa$ of the interface in the cells nearest to the rod such that 
\begin{equation}
\boldsymbol{n} = \frac{\boldsymbol{\nabla} \alpha_{w}}{|\boldsymbol{\nabla}\alpha_{w}|}		
\end{equation}
\begin{equation}
\kappa = \boldsymbol{\nabla.n}
\end{equation}
The surface tension force term is evaluated as follows
\begin{equation}
\boldsymbol{F_{\sigma}} = \sigma_{a}(\frac{\kappa \boldsymbol{\nabla} \alpha_{w} \rho}{\frac{\rho_{w} + \rho_{o}}{2}})
\end{equation}
The effect of the contact angle thus gets incorporated into the dynamics of the interface through the change in curvature and hence $\boldsymbol{F_{\sigma}}$. The $\boldsymbol{F_{\sigma}}$ term captures the force due to the surface tension and It is mathematically modeled by the Continuum Surface Force model \cite{brackbill1992continuum}. The aforementioned equations are dependent on the volume fractions of each fluid within the cell. Further details pertaining to the simulations are provided in the supplementary note S1.

\section{Results and Discussion}
\begin{figure*}
\centering
\includegraphics[width=1\linewidth]{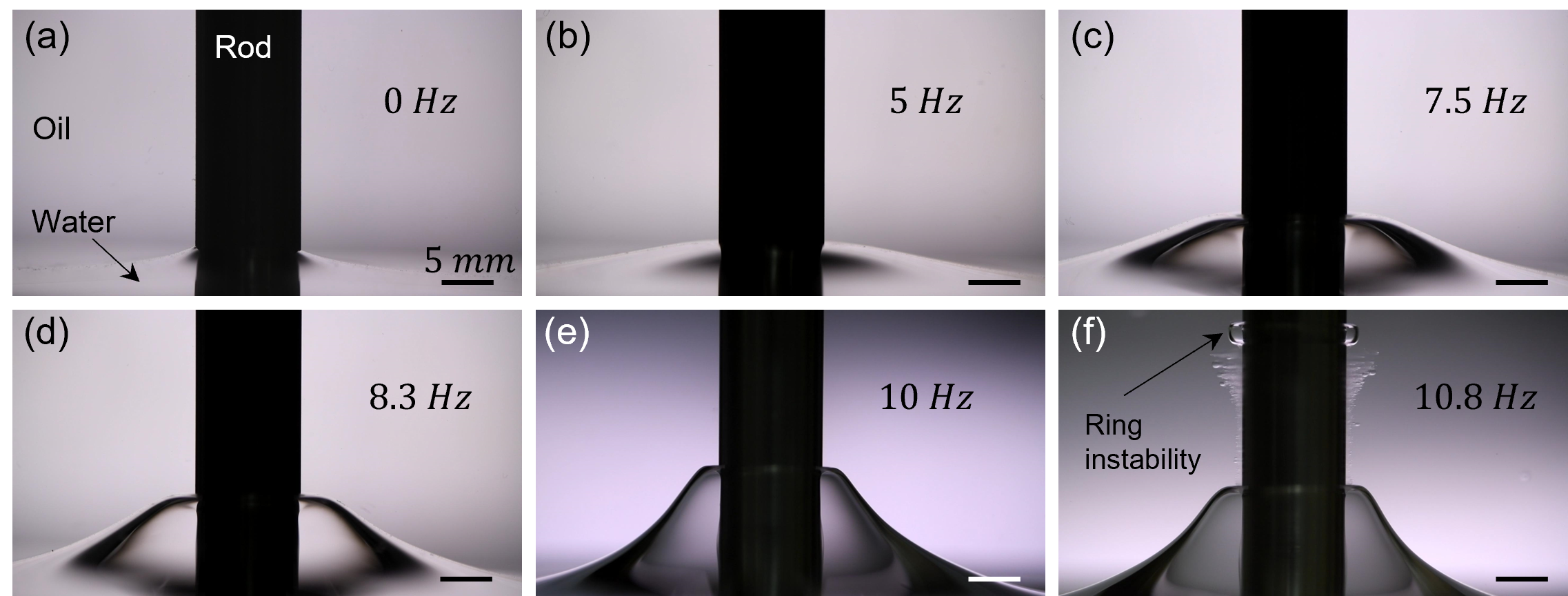}
\caption{(a)-(e) Steady state profile of oil-water interface near the rod rotating at 0, 5, 7.5, 8.3 and 10 Hz respectively. (e) Emulsification and ring instability at $\omega$=10.8 Hz.}
\label{fig:general_RC}
\end{figure*}
\begin{figure*}
\centering
\includegraphics[width=0.65\linewidth]{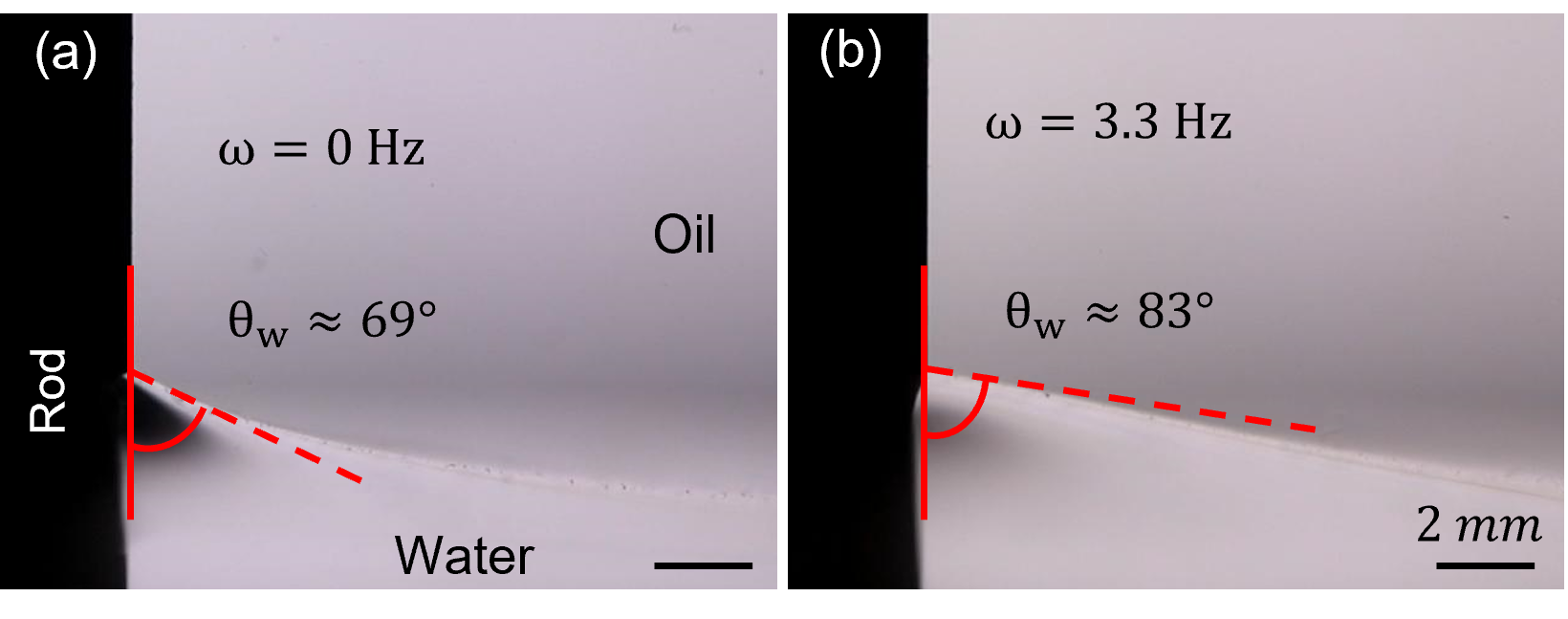}
\caption{Right half of the stable interfacial profile near the rod showing the presence of CAH at the oil-water-rod interface. CAH can be observed by comparing the contact angle, $\theta_w$ for (a) static rod, and (b) rod rotating at 3.3 Hz.}
\label{fig:CAH}
\end{figure*}
In the present setup, as the rod is set to rotate at a constant speed, the oil-water interface achieves a stable steady state profile after few seconds of initial transience. Figure \ref{fig:general_RC}(a-e) presents the steady state interface shape for different rod rotation speed. A stable shape of interfacial profile is observed only below a maximum rod rotation speed, $\omega_{max}$ (10 Hz for the present case). Beyond this speed, instead of a stable interface, water-in-oil emulsion is formed through sheet breakup and ring instability as shown in Figure \ref{fig:general_RC}f and supplementary movie S1 (detailed discussion in the later section). The present study is mainly focused on the regime $\omega \leq \omega_{max}$ where stable interfacial profile is observed.

\begin{figure}
\centering
\includegraphics[width=0.7\linewidth]{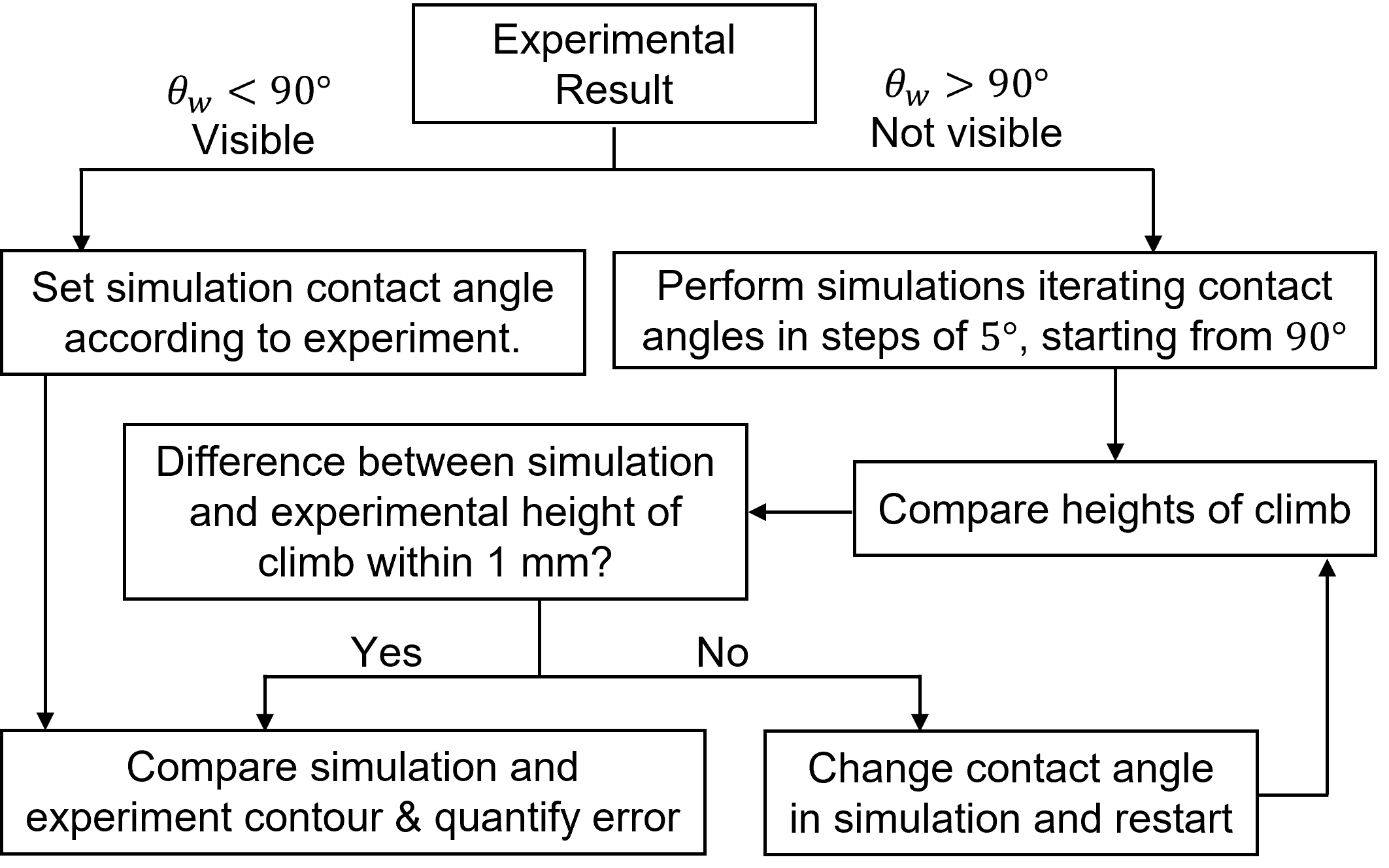}
\caption{Flowchart illustrating the methodology of present work.}
\label{fig:flow_chart}
\end{figure}

The existing literature on the rod climbing of Newtonian liquids does not elaborate on the effects of contact angle and assumes that it has a constant value of 90$^{\circ}$. Recently \citet{chandra2021contact} have shown that the CAH due to the pinned CL plays an important role in modulating the rod climbing effect of viscoelastic liquids. Similar observations are made in the present study for the rod climbing of Newtonian liquids. The contact angle $\theta_w$, at the three-phase CL of oil-water-rod interface is neither constant nor necessarily 90$^{\circ}$ as shown in Figure \ref{fig:CAH}. For the materials considered in the present condition $\theta_w\approx $ 69$^{\circ}$ when the rod is in static condition (Figure \ref{fig:CAH}a). As the rod is set to rotate at some constant frequency, the oil-water interface tries to go upwards to balance the inertial force due to secondary flows in the lower liquid \citep{bonn2004rod}. Here, we observed that this inertial force is manifested not only in terms of vertical displacement of oil-water interface, but also in terms of the change in $\theta_w$ (Figure \ref{fig:CAH}b).
\begin{figure*}
\centering
\includegraphics[width=0.8\linewidth]{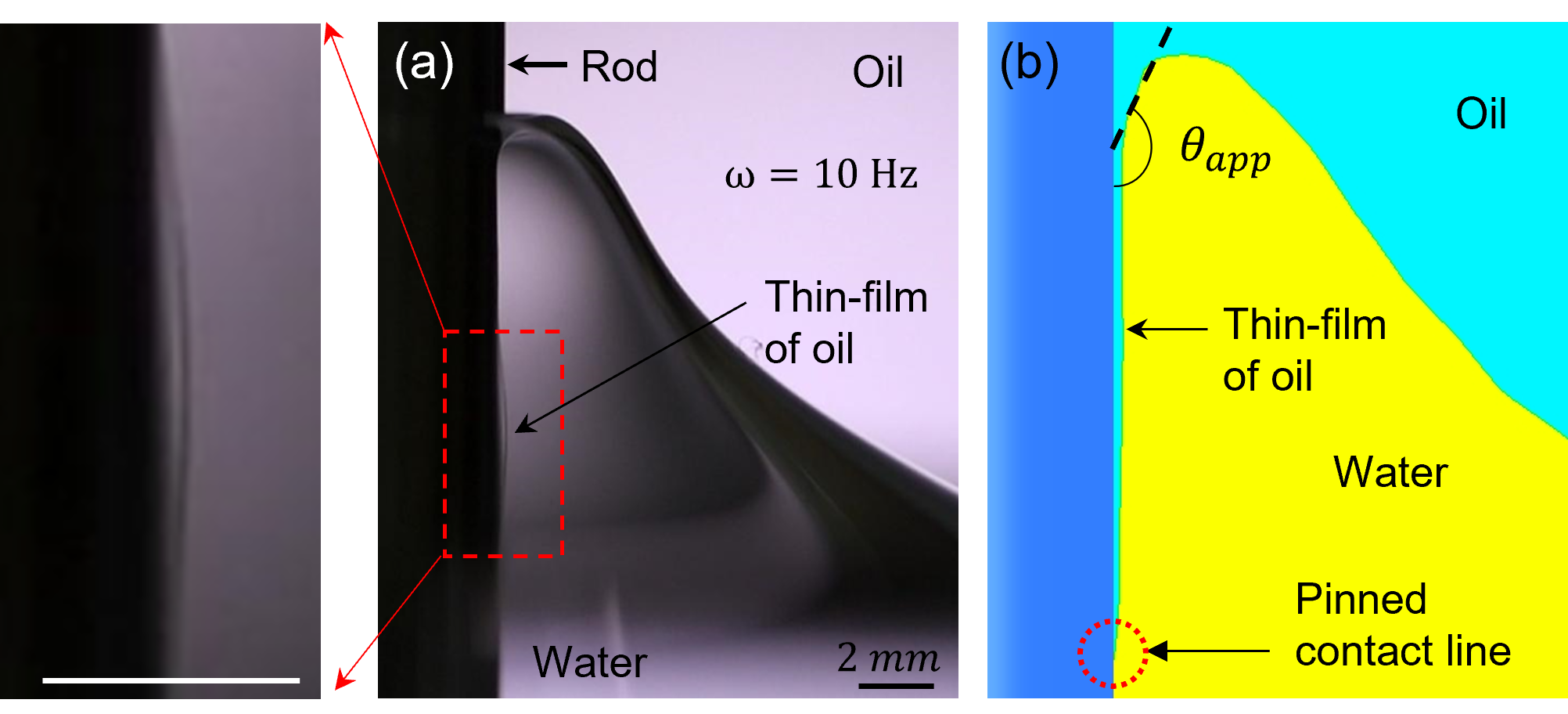}
\caption{Topology of the steady state interface in the vicinity of the rod rotating at high frequencies ($\omega>$5 Hz). (a) Experimental image showing thin film of oil trapped between water and the rod surface at $\omega=$10 Hz. (b) Schematic showing macroscopic contact angle $\theta_{app}$ at the location of apparent contact between the rod and the oil-water interface, if the thin oil film is neglected.}
\label{fig:theta_app_oil_film}
\end{figure*}
There is a gradual increase of $\theta_w$ with increasing speed of rod rotation. For $\omega$ up to 3.3 Hz, $\theta_w$ is less than 90$^{\circ}$, and it is experimentally observable as shown in Figure \ref{fig:CAH}. With further increase in $\omega$,  $\theta_w$  becomes greater than 90$^{\circ}$. For the case of $\theta_w>$ 90$^{\circ}$, the three-phase CL is hidden behind the deformed interface, and it cannot be imaged due to the lensing action of the bulged liquid meniscus. Experiments at even higher frequencies reveal the existence of a very thin film of oil between the climbed interface and the rod as shown in Figure \ref{fig:theta_app_oil_film}. This suggests the possibility that the three-phase CL always remains pinned at its initial position (refer the schematic in Figure \ref{fig:theta_app_oil_film}b) throughout the course of the rod climbing experiments, and the observed climb occurs in a region of liquid slightly away from the rod surface. Hints for this possibility is also evident from the motion of the receding meniscus just after the rotation is stopped (refer to the supplementary movie S2). As the rod is stopped after rotating it at a sufficiently high speed, the climbed interface starts receding and reveals a thin layer of oil trapped between the rod and the climbed portion of water. In such a scenario (Figure \ref{fig:theta_app_oil_film}a), it becomes difficult to define and measure the contact angle formed at the actual location of the three-phase CL. Therefore, we define an apparent contact angle, $\theta_{app}$ which is the angle that the oil-water interface appears to make with the rod just before experiencing an almost singular dip (Figure \ref{fig:theta_app_oil_film}b). $\theta_{app}$ can also be thought as the contact angle that the oil-water interface appears to make with the rod surface if the thin film of oil is neglected. $\theta_{app}$ is used in place of $\theta_w$ in the simulations for the cases of the rod angular frequencies where $\theta_w\geq$ 90$^{\circ}$ in the experiments. The detailed methodology adopted in the present work for supplementing the experimental results with simulations is illustrated in the form of a flowchart in Figure \ref{fig:flow_chart}. Process is straightforward for lower rod rotation frequencies where $\theta_w<$ 90$^{\circ}$. In these cases, the input value of $\theta_w$ in simulations is directly obtained from the experimental images. For the cases where $\theta_w\geq$ 90$^{\circ}$, we use the apparent contact angle, $\theta_{app}$ in place of $\theta_w$. $\theta_{app}$ cannot be measured from the experimental images due to the hindrance of the bulged meniscus. Therefore, it has been used as a fitting parameter in the numerical simulations, and its value is iterated in steps of 5$^{\circ}$ starting from 90$^{\circ}$ to match the climbing height obtained from the corresponding experiment. Once the suitable $\theta_{app}$ is known, contour of simulated interfacial profile is compared with experiments and error is estimated.
\begin{figure}
\centering
\includegraphics[width=0.6\linewidth]{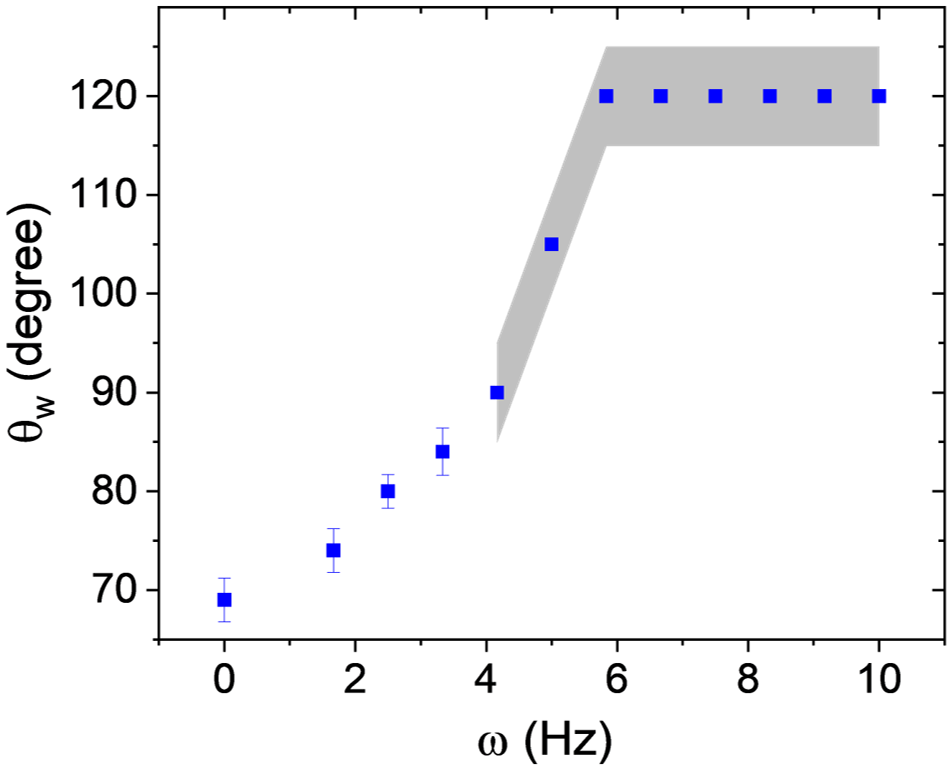}
\caption{Variation of the input contact angle with the angular frequency of the rod. Standard error in case of $\theta_{w} < 90^{o}$ is obtained from the experiments. For $\theta_{w} \geq 90^{o}$ there is an uncertainty of $5^{o}$ represented by the shaded region.}
\label{fig:fig_theta_vs_omega_plot.png}
\end{figure}
The values of input contact angle at different $\omega$ as used in numerical simulations are shown in Figure \ref{fig:fig_theta_vs_omega_plot.png}. For $\omega\leq$3.3 Hz, the value of $\theta_w$ is less than 90$^{\circ}$ and the presented values have been obtained directly from the measurements using the experimental images. The error bar represents the standard error of three repeated experimental runs. For $\omega>$3.3 Hz, the value of $\theta_w$ goes beyond 90$^{\circ}$ which cannot be measured experimentally therefore $\theta_{app}$ has been used in place of $\theta_w$. Since the simulation are performed by iterating the $\theta_{app}$ in steps of 5$^{\circ}$, hence the gray shaded region (Figure \ref{fig:fig_theta_vs_omega_plot.png}) represents the region of uncertainty.

\begin{figure*}
\centering
\includegraphics[width=0.9\linewidth]{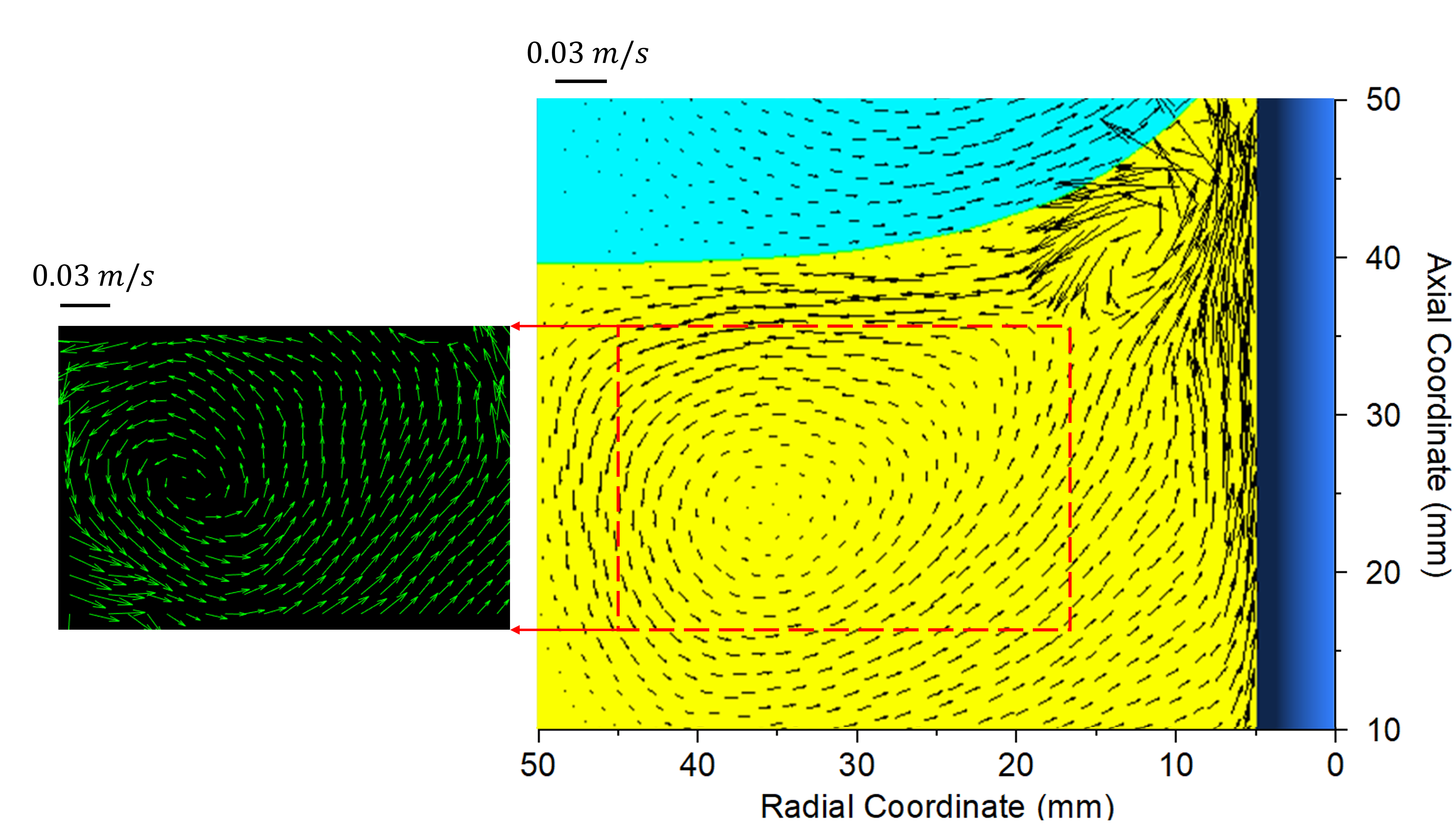}
\caption{Velocity field in the meridional plane of the water phase obtained from experiments using PIV (left) and from simulations (right) at $\omega=10 Hz$ for the geometry used in the present work. Red box indicates the region of interest for PIV.}
\label{fig:PIV}
\end{figure*}

It is worth noting that in the process of defining $\theta_{app}$, the thin film of oil between the rod surface and the oil-water interface is neglected. This thin film formation is purely an interfacial phenomenon which occurs due to the interplay of the energies among the water-oil, water-rod, and oil-rod interfaces \citep{joseph_nguyen_beavers_1984, joseph_renardy_renardy_nguyen_1985}. Therefore, even if the film is neglected in the numerical simulation, there is no significant change in the hydrodynamics of the problem. To probe this further, velocity field of secondary flows in meridional plane of water phase is measured using particle image velocimetry (PIV) technique. This showed a satisfactory match with the flow field obtained from the numerical simulations as shown in Figure \ref{fig:PIV}. Here, presence of the bulged meniscus, glare from the rod surface due to laser illumination, and limited field of view did not allow us to perform the PIV measurements near the rod. Therefore, PIV data is shown (Figure \ref{fig:PIV}) only for a region away from the rod. The match of secondary flow field between experiment and simulation supports our hypothesis that the main physics of the Newtonian rod climbing remains unaffected even if the thin oil film is neglected. The effect of this thin film is indirectly captured in the simulations by virtue of the $\theta_{app}$ which was implicitly kept as 90$^{\circ}$ in the simulations of \citet{zhao2017deformation}.

\begin{figure*}
\centering
\includegraphics[width=0.8\linewidth]{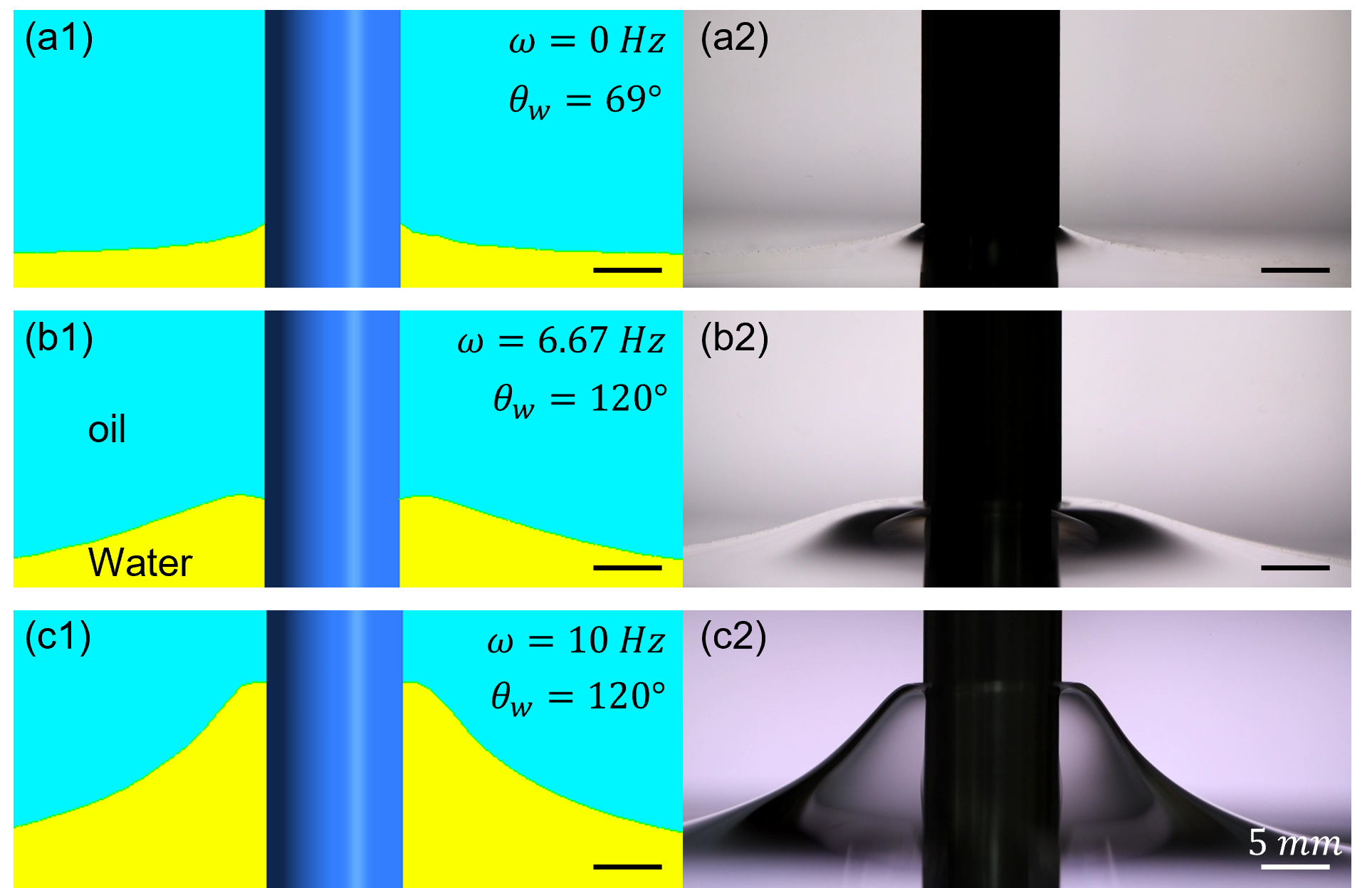}
\caption{Stable interfacial profile obtained from simulations (a1,b1,c1) and corresponding experiments (a2,b2,c2) for rod rotation speeds of 0, 6.67 and 10 Hz respectively.}
\label{fig:Sim_exp}
\end{figure*}

Present study shows that $\theta_w$ plays an important role in the rod climbing effect of Newtonian liquids. By providing the appropriate values of $\theta_w$ (from Figure \ref{fig:fig_theta_vs_omega_plot.png}) as a boundary condition, different aspects of the Newtonian rod climbing effect can be captured properly in the numerical simulations. A visual comparison of simulated and experimental interfacial profiles for different rod rotation speeds are shown in Figure \ref{fig:Sim_exp}.
Present work investigates the effect of $\theta_w$ in terms of the climbing height, the threshold rod rotation speed for onset of climbing, and the interfacial profile. Discussion on each aspect is provided in the following subsection.

\subsection{Climbing height and threshold rod rotation speed}
\begin{figure*}
\centering
\includegraphics[width=0.8\linewidth]{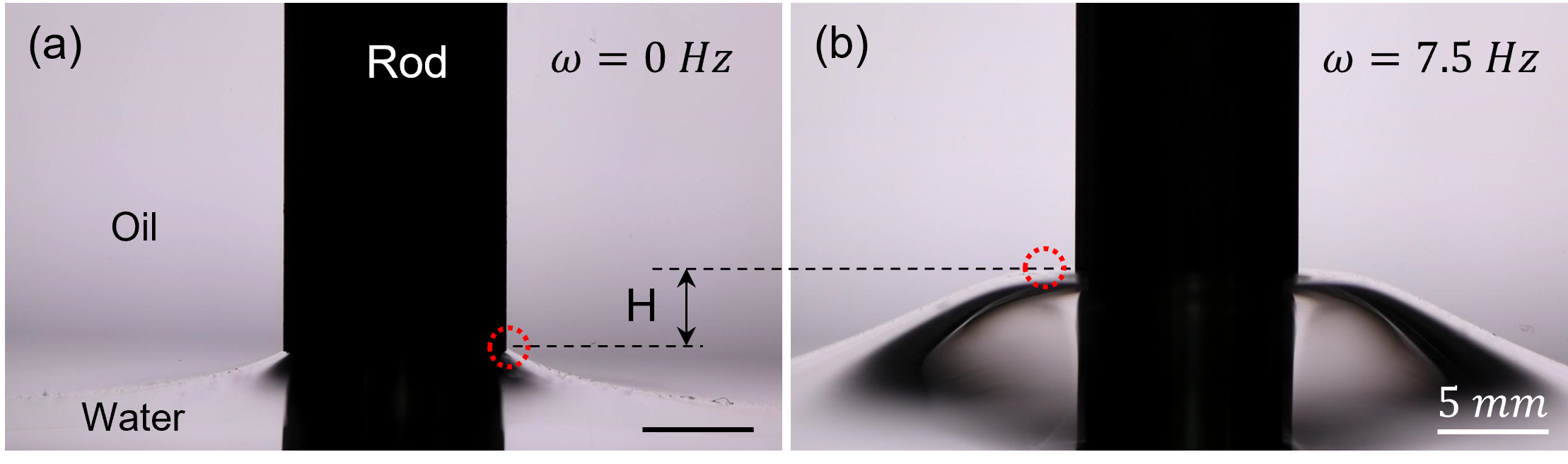}
\caption{The stable interfacial profile when the rod is (a) in static condition and (b) rotating at $\omega = $7.5 Hz. The height of climb of the interface is denoted by H. Red circle indicates the maximum elevation point of the interfacial profile.}
\label{fig:Climbing_height_def}
\end{figure*}

It is already shown that the climbed interface experiences a sharp dip extremely close to the rod (Figure \ref{fig:theta_app_oil_film}) therefore, climbing height cannot be defined in terms of the vertical displacement of the three-phase CL on the rod surface as done in our previous work \citep{chandra2021contact}. Here we define the climbing height, $H$ as the vertical distance between the points of maximum elevation of the steady state interface at a given $\omega$ and the static condition as shown in Figure \ref{fig:Climbing_height_def}.
\begin{figure}
\centering
\includegraphics[width=0.6\linewidth]{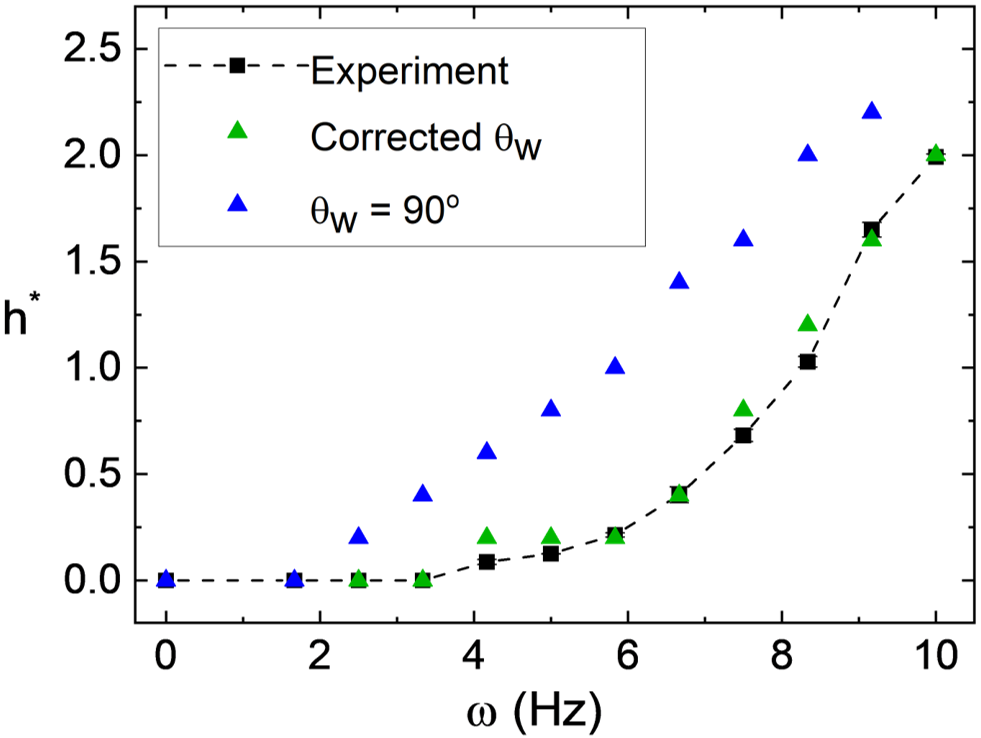}
\caption{Variation of the simulation and experimental non-dimensional climbing height, $h^{*} = \frac{H}{R}$ with the angular frequency of the rod.}
\label{fig:climbing_height_plot}
\end{figure}
The climbing heights at different $\omega$ obtained from experiments and simulation are shown in Figure \ref{fig:climbing_height_plot}. Simulation results are shown for two different cases- first, assuming a flat meniscus near the rod surface, i.e., $\theta_w=$ 90$^{\circ}$ as used in the existing literature \citep{zhao2017deformation}. Second, by using the corrected value of $\theta_w$ (from Figure \ref{fig:fig_theta_vs_omega_plot.png}) as proposed in the present study. It is clear from Figure \ref{fig:climbing_height_plot} that, if a constant value of $\theta_w=$90$^{\circ}$ is considered i.e., the CAH is ignored then the climbing height obtained from simulation deviates significantly from the experimental values. 
Also, the threshold rod rotation speed, $\omega_{th}$ for non-zero climbing is predicted to lie between 1.67 to 2.5 Hz. Whereas zero climbing is observed even at 3.33 Hz in the experiments. Accounting for the CAH and hence by using the appropriate value of $\theta_w$, the climbing height as well as $\omega_{th}$ is predicted properly in the simulations. In the existing literature, $\omega_{th}$ is attributed to the critical rod rotation speed for the onset of secondary flows due to Taylor-Couette instability in the heavier liquid \citep{bonn2004rod}. The present study shows that $\omega_{th}$ is the rod rotation speed required not just for the onset but to produce the secondary flows strong enough to either overcome the pinning force or to increase the $\theta_w$ beyond 90$^{\circ}$, whichever is achieved earlier. It should be noted that, for $\omega>$ 3.3 Hz, the value of $\theta_w$ goes beyond 90$^{\circ}$ and $\theta_w=\theta_{app}$ is chosen in such a way that the simulated climbing height matches with the experiments. In such cases the veracity of the proposed mechanism is ascertained by looking at the match of interfacial profiles obtained from simulation and experiments.

\subsection{Interfacial profile}
\begin{figure*}
\centering
\includegraphics[width=1\linewidth]{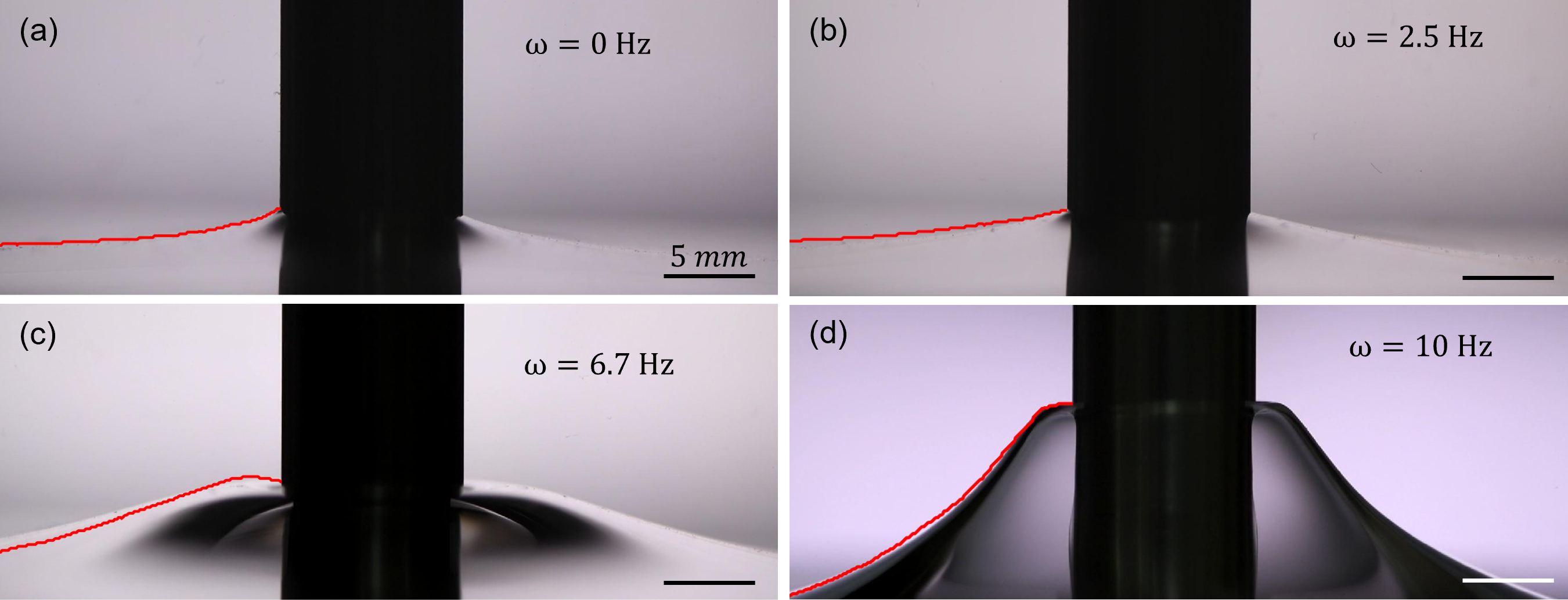}
\caption{Match of the steady state interfacial profile obtained from experiments and simulation (red contour) at different rod-rotation frequencies: (a) 0 Hz (b) 2.5 Hz (c) 6.7 Hz (d) 10 Hz.}
\label{fig:contour_match}
\end{figure*}
Figure \ref{fig:contour_match} presents the interfacial profile obtained from simulation (red curve) overlapped on the experimental images. For $\omega\leq$3.33 Hz the value of $\theta_w$ is less than 90$^{\circ}$ which is measured directly from the experimental images. Providing these measured values of $\theta_w$ as input to the simulations, a proper prediction of climbing height (Figure \ref{fig:Climbing_height_def}) and interfacial profile is obtained (Figure \ref{fig:contour_match}a and \ref{fig:contour_match}b). For $\omega>$3.33 Hz, the value of $\theta_w=\theta_{app}$ is chosen to match simulation climbing height with the experiments. For the same value of $\theta_{app}$ a good agreement is found between the experiments and the simulated interfacial profile (Figure \ref{fig:contour_match}c and \ref{fig:contour_match}d). It is worth noting that the contact angle obtained from the output of the simulation is slightly different from the value provided as an input (120$^{\circ}$ for Figure \ref{fig:contour_match}c and \ref{fig:contour_match}d). The reason for this lies in the linear interpolation of the geometric reconstruction scheme to reconstruct the interface (refer to supplementary information Note S2). The error in the match between the experimental and the simulated interfacial profile is within a root mean square value of 0.6 mm for all $\omega$ considered in the present study. This error is calculated by taking the root mean square of the vertical deviation between simulation and experimental interfaces at 1000 equally spaced radial locations spanning from the rod surface to 4.5$R$ ($R$ being the rod radius). Details of error estimation procedure is provided in the supplementary document (Note S3).
It is clear from Figure \ref{fig:climbing_height_plot} and Figure \ref{fig:contour_match} that the modification of $\theta_w$ captures the interface deformation both qualitatively and quantitatively. This reinforces the fact that, neglecting the thin oil film present between the deformed interface and the rod, does not lead to any loss of the physics of the problem if $\theta_w$ is properly accounted for.
\begin{figure}
\centering
\includegraphics[width=0.7\linewidth]{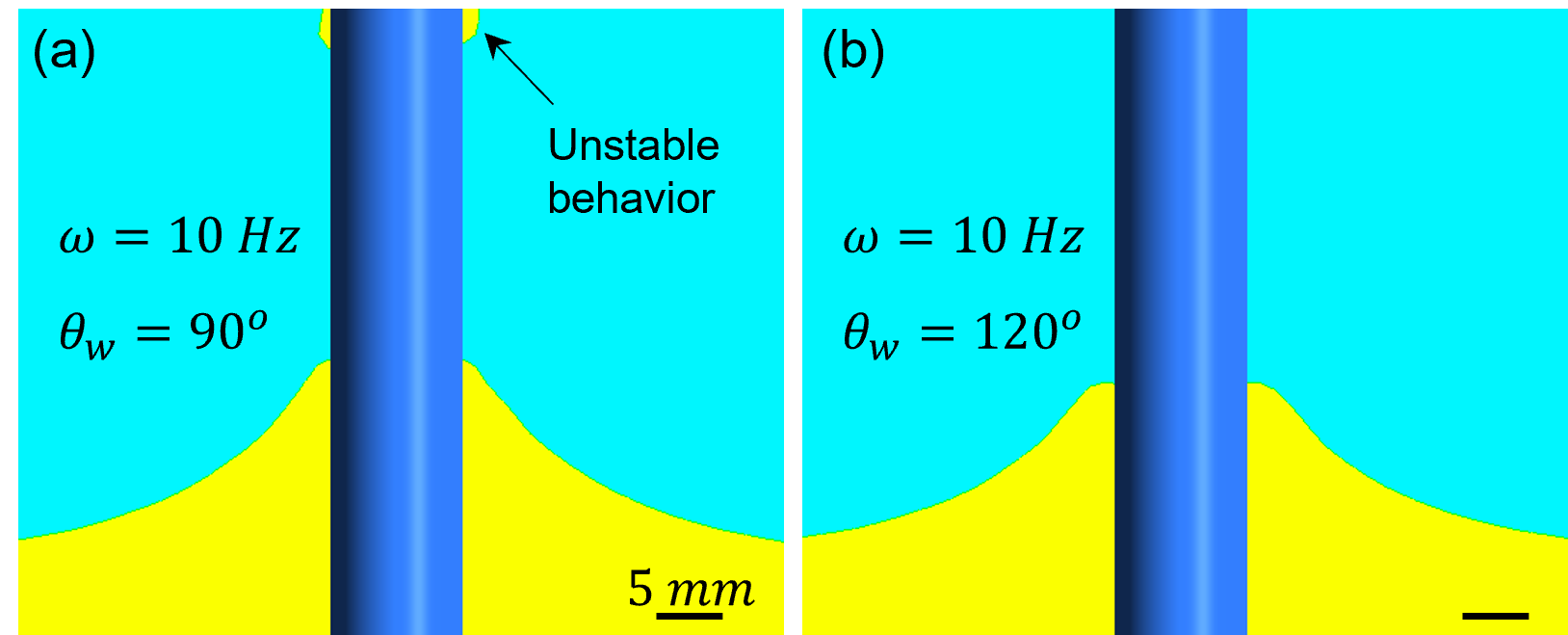}
\caption{Configuration of the two liquids obtained from simulation at $\omega$=10Hz for (a)$\theta_w$=$90^{\circ}$ (b)$\theta_w$=$120^{\circ}$. Interface shows unstable behavior in the experimental time scale for $\theta_w$=$90^{\circ}$, and the presented figure corresponds to the state after running the simulation for 20 seconds of physical time. Stable interface is obtained for $\theta_w$=$120^{\circ}$.  }
\label{fig:instability_sim}
\end{figure}
Another interesting observation is that the simulation predicts an unstable interface at $\omega$=10 Hz for $\theta_w$=90$^{\circ}$ (Figure \ref{fig:instability_sim}a). Whereas, $\theta_w$=120$^{\circ}$ predicts stable interface at the same rotation frequency (Figure \ref{fig:instability_sim}b) which matches perfectly with the experimental observation (Figure \ref{fig:contour_match}d). This further supports our claim that the contact angle must be corrected to properly model the Newtonian rod climbing effect.

\subsection{Trend of $\theta_w$ with $\omega$}
\begin{figure}
\centering
\includegraphics[width=0.6\linewidth]{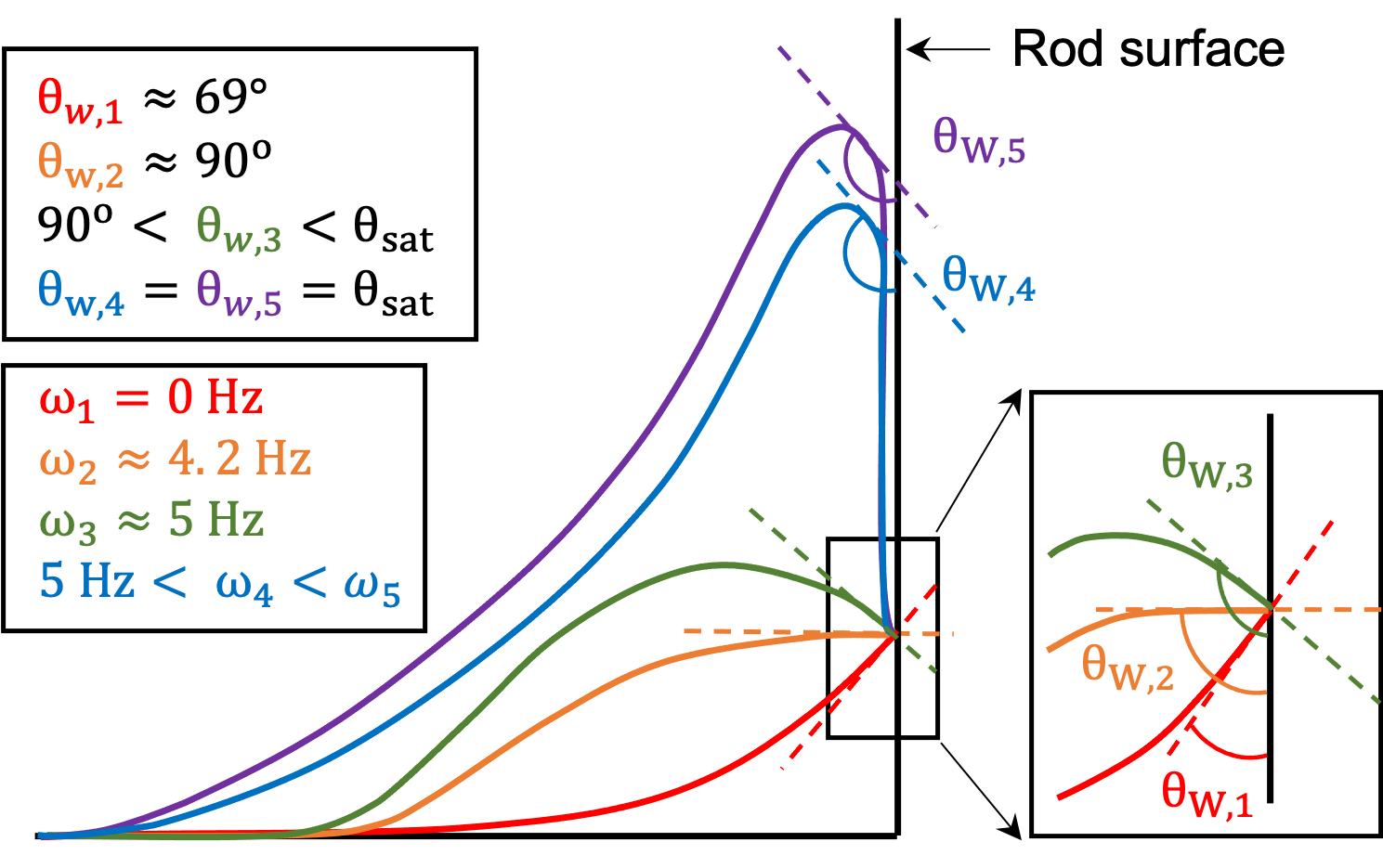}
\caption{Schematic showing the interfacial profile and the corresponding contact angle at different rod rotation frequencies. This depicts the mechanism for saturation of contact angle beyond a certain frequency.}
\label{fig:theta_omega_schematic}
\end{figure}
It is observed that the variation of $\theta_{w}$  with $\omega$ shows an interesting trend (Figure \ref{fig:fig_theta_vs_omega_plot.png}). There is an initial increment and then saturation of $\theta_{w}$ with increasing $\omega$, and this can be explained by accounting for the observed thin film of oil. Figure \ref{fig:theta_omega_schematic} shows the schematic of interfacial profile corresponding to different rod rotation speed, $\omega_i$ and respective contact angle, $\theta_{w,i}$. At frequencies lower than $\omega_{th}$ ($\sim$ 4.2 Hz for the present case) there is a gradual change in $\theta_{w}$ which happens because the three-phase CL is pinned on the rod surface. However, once the rod rotation speed exceeds $\omega_{th}$ there is a very steep increase in $\theta_{w}$ until it saturates at a particular value, $\theta_{sat}$ which in case of the present study is 120$^{\circ}$  with an uncertainty of $\pm$5$^{\circ}$. At such high frequencies, the system occupies a configuration where a thin film of oil is present between the rod and the climbed meniscus (Figure \ref{fig:theta_app_oil_film}). In such case, $\theta_{w}$ is replaced by $\theta_{app}$ which is the angle formed at the location of apparent contact between the rod and water-oil interface. Since there is no actual contact of the interface with the rod surface, pinning and hence CAH cannot occur. This is analogous to the de-pinned CL on the oil coated rod surface of our previous work related to the rod climbing effect of viscoelastic liquids \citep{chandra2021contact}. This explains that the contact angle will not change further and assumes a constant saturated value.

\subsection{Discussion on ring instability}
\begin{figure*}
\centering
\includegraphics[width=1\linewidth]{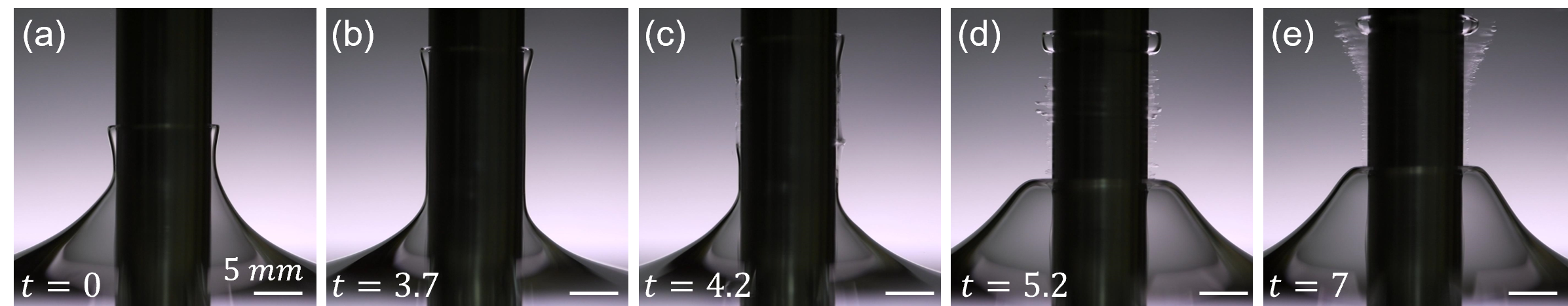}
\caption{Transient evolution of interfacial profile at $\omega = 10.8$ Hz. Time is in seconds. The frame at $t = 0$ s corresponds to the initiation of an axisymmetric finger-like structure. (a) - (b) This structure begins to move along the rod surface at an average velocity of 2 mm/s. (c) At $t = 4.2$ s the sheet of water breaks down. It gets fully emulsified within a second as can be observed in (d). This causes the finger-like structure to get disconnected from the bulk fluid, thereby showing a ring-instability (d) and (e).}
\label{fig:ring_instability}
\end{figure*}

As the $\omega$ is increased beyond a certain value, $\omega_{max}$ (10 Hz for the present case), a steady shape of interface is not observed. Formation of ring instability and water-in-oil emulsion happens at such high rotation frequencies \citep{bonn2004rod}. Figure \ref{fig:ring_instability} presents the transient evolution of interfacial profile and subsequent emulsification at $\omega=$10.8 Hz. The process of emulsification begins with the formation of a finger-like structure (Figure \ref{fig:ring_instability}a), which starts moving up along the rod with a velocity $\sim$2 mm/s for roughly 3.7 s. This causes the formation of an enclosing sheet of water, referred as sheet instability by \citet{zhao2017deformation}, around the inner thin film of oil (Figure \ref{fig:ring_instability}b). Once the finger-like structure has reached its maximum height, the water sheet persists for $\sim$0.5 s before it breaks up and gets emulsified (Figure \ref{fig:ring_instability}c). This causes the uppermost finger-like structure to get disconnected from the bulk of the water and form a ring-like structure around the rod (Figure \ref{fig:ring_instability}d and \ref{fig:ring_instability}e). This is the ring-instability as reported by \citet{bonn2004rod}. The ring also gets emulsified in few seconds and the emulsion gets dispersed into the bulk of the oil. After this, another finger-like structure forms at the interface and the entire process repeats itself. It is observed that every successive cycle of ring instability and emulsification takes a longer period of time than the previous one (refer to the supplementary movie S1). In the first cycle the water sheet persisted for $\sim$0.5 s while in the third cycle it persisted for more than 40 s. The possible explanation for this observation may lie in the altered hydrodynamic conditions due to increased amount of water-in-oil emulsion, and studying this is beyond the scope of the present work. 
The method of correcting $\theta_w$ in simulations to properly capture the rod climbing effect of Newtonian liquids as proposed in the present study is applicable only for the regime $\omega\leq\omega_{max}$ where steady state profile is observed. In this regime, the thin film of oil can be neglected, and the physics can be captured through an apparent contact angle. However, if $\omega>\omega_{max}$ then the simulation with the present parameter cannot be used to supplement the experimental results because the interface no longer reaches a steady state but shows a highly transient and unstable behavior (Figure \ref{fig:ring_instability}). Moreover, the $\omega>\omega_{max}$ regime involves emulsification through the breakup of a thin film. This would require a very fine mesh in the simulation to accurately capture this phenomena, such that it can also account for the thin film of oil and water. Hence, we limit the numerical results of the present study till the regime of angular frequencies $\omega \leq \omega_{max}$ where a steady state interface exists, and the thin film of oil can be neglected.

\section{Conclusion}
In the present work we have studied the rod climbing effect of Newtonian liquids in a stratified system with Silicone oil of viscosity 355 mPa-s being the lighter liquid and DI water as the heavier liquid. Using a stainless steel rod of diameter 10 mm, steady state climbing profiles are observed for rotation speeds, $\omega$ up to 10 Hz. Beyond this speed, ring instability and subsequent emulsification is observed. Experiments revealed that the contact angle $\theta_{w}$ at the oil-water-rod interface shows a hysteresis based on the rod rotation speed. To precisely model the Newtonian rod climbing phenomenon, we propose that the contact angle hysteresis must be accounted in the boundary condition by using the correct value of $\theta_w$ at any given $\omega$. The proposed methodology yields a quantitative match between experiments and simulations in terms of climbing height, threshold rod rotation speed, and shape of the oil-water interface. For higher values of $\omega$ ($>$5Hz), the value of $\theta_w$ becomes more than 90$^{\circ}$ and cannot be measured experimentally due to obstruction from the bulged meniscus. At such high frequencies of rod rotation a thin film of oil trapped between the rod and the climbed portion of water is also observed. In these cases, the thin film of oil can be neglected and the dynamics of Newtonian rod climbing can be captured in terms of an apparent contact angle, $\theta_{app}$. Using it as a fitting parameter, the value of $\theta_{app}$ is obtained from the simulation to match the experimental climbing height. Reliability of $\theta_{app}$ is confirmed by matching the simulated interfacial profile with the experiments. The proposed methodology is applicable only for the $\omega$ up to which a steady and stable interface is observed (10 Hz for the present case). Beyond this $\omega$, the proposed method cannot be applied due to the transient nature of the phenomenon, and  requirement of a very fine grid size in simulations to account for emulsification assisted by thin film breakup.

\section*{Data Availability Statement}
The data that support the findings of this study are available within the article and its supplementary material. A supplementary document, and two supplementary movies are provided as supplementary material for this article.

\nocite{*}
\bibliography{aipsamp}

\providecommand{\noopsort}[1]{}\providecommand{\singleletter}[1]{#1}%
\begin{thebibliography}{18}%
\makeatletter
\providecommand \@ifxundefined [1]{%
 \@ifx{#1\undefined}
}%
\providecommand \@ifnum [1]{%
 \ifnum #1\expandafter \@firstoftwo
 \else \expandafter \@secondoftwo
 \fi
}%
\providecommand \@ifx [1]{%
 \ifx #1\expandafter \@firstoftwo
 \else \expandafter \@secondoftwo
 \fi
}%
\providecommand \natexlab [1]{#1}%
\providecommand \enquote  [1]{``#1''}%
\providecommand \bibnamefont  [1]{#1}%
\providecommand \bibfnamefont [1]{#1}%
\providecommand \citenamefont [1]{#1}%
\providecommand \href@noop [0]{\@secondoftwo}%
\providecommand \href [0]{\begingroup \@sanitize@url \@href}%
\providecommand \@href[1]{\@@startlink{#1}\@@href}%
\providecommand \@@href[1]{\endgroup#1\@@endlink}%
\providecommand \@sanitize@url [0]{\catcode `\\12\catcode `\$12\catcode
  `\&12\catcode `\#12\catcode `\^12\catcode `\_12\catcode `\%12\relax}%
\providecommand \@@startlink[1]{}%
\providecommand \@@endlink[0]{}%
\providecommand \url  [0]{\begingroup\@sanitize@url \@url }%
\providecommand \@url [1]{\endgroup\@href {#1}{\urlprefix }}%
\providecommand \urlprefix  [0]{URL }%
\providecommand \Eprint [0]{\href }%
\providecommand \doibase [0]{http://dx.doi.org/}%
\providecommand \selectlanguage [0]{\@gobble}%
\providecommand \bibinfo  [0]{\@secondoftwo}%
\providecommand \bibfield  [0]{\@secondoftwo}%
\providecommand \translation [1]{[#1]}%
\providecommand \BibitemOpen [0]{}%
\providecommand \bibitemStop [0]{}%
\providecommand \bibitemNoStop [0]{.\EOS\space}%
\providecommand \EOS [0]{\spacefactor3000\relax}%
\providecommand \BibitemShut  [1]{\csname bibitem#1\endcsname}%
\let\auto@bib@innerbib\@empty
\bibitem [{\citenamefont {Bonn}\ \emph {et~al.}(2004)\citenamefont {Bonn},
  \citenamefont {Kobylko}, \citenamefont {Bohn}, \citenamefont {Meunier},
  \citenamefont {Morozov},\ and\ \citenamefont {Van~Saarloos}}]{bonn2004rod}%
  \BibitemOpen
  \bibfield  {author} {\bibinfo {author} {\bibfnamefont {D.}~\bibnamefont
  {Bonn}}, \bibinfo {author} {\bibfnamefont {M.}~\bibnamefont {Kobylko}},
  \bibinfo {author} {\bibfnamefont {S.}~\bibnamefont {Bohn}}, \bibinfo {author}
  {\bibfnamefont {J.}~\bibnamefont {Meunier}}, \bibinfo {author} {\bibfnamefont
  {A.}~\bibnamefont {Morozov}}, \ and\ \bibinfo {author} {\bibfnamefont
  {W.}~\bibnamefont {Van~Saarloos}},\ }\bibfield  {title} {\enquote {\bibinfo
  {title} {Rod-climbing effect in newtonian fluids},}\ }\href@noop {}
  {\bibfield  {journal} {\bibinfo  {journal} {Physical review letters}\
  }\textbf {\bibinfo {volume} {93}},\ \bibinfo {pages} {214503} (\bibinfo
  {year} {2004})}\BibitemShut {NoStop}%
\bibitem [{\citenamefont {Zhao}\ \emph {et~al.}(2017)\citenamefont {Zhao},
  \citenamefont {Gentric}, \citenamefont {Dietrich}, \citenamefont {Ma},\ and\
  \citenamefont {Li}}]{zhao2017deformation}%
  \BibitemOpen
  \bibfield  {author} {\bibinfo {author} {\bibfnamefont {C.}~\bibnamefont
  {Zhao}}, \bibinfo {author} {\bibfnamefont {C.}~\bibnamefont {Gentric}},
  \bibinfo {author} {\bibfnamefont {N.}~\bibnamefont {Dietrich}}, \bibinfo
  {author} {\bibfnamefont {Y.}~\bibnamefont {Ma}}, \ and\ \bibinfo {author}
  {\bibfnamefont {H.~Z.}\ \bibnamefont {Li}},\ }\bibfield  {title} {\enquote
  {\bibinfo {title} {Deformation of liquid-liquid interfaces by a rotating
  rod},}\ }\href@noop {} {\bibfield  {journal} {\bibinfo  {journal} {Physics of
  Fluids}\ }\textbf {\bibinfo {volume} {29}},\ \bibinfo {pages} {072108}
  (\bibinfo {year} {2017})}\BibitemShut {NoStop}%
\bibitem [{\citenamefont {Dealy}\ and\ \citenamefont
  {Vu}(1977)}]{dealy1977weissenberg}%
  \BibitemOpen
  \bibfield  {author} {\bibinfo {author} {\bibfnamefont {J.}~\bibnamefont
  {Dealy}}\ and\ \bibinfo {author} {\bibfnamefont {T.}~\bibnamefont {Vu}},\
  }\bibfield  {title} {\enquote {\bibinfo {title} {The weissenberg effect in
  molten polymers},}\ }\href@noop {} {\bibfield  {journal} {\bibinfo  {journal}
  {Journal of Non-Newtonian Fluid Mechanics}\ }\textbf {\bibinfo {volume}
  {3}},\ \bibinfo {pages} {127--140} (\bibinfo {year} {1977})}\BibitemShut
  {NoStop}%
\bibitem [{\citenamefont {Weissenberg}(1947)}]{weissenberg1947continuum}%
  \BibitemOpen
  \bibfield  {author} {\bibinfo {author} {\bibfnamefont {K.}~\bibnamefont
  {Weissenberg}},\ }\bibfield  {title} {\enquote {\bibinfo {title} {A continuum
  theory of rhelogical phenomena},}\ }\href@noop {} {\  (\bibinfo {year}
  {1947})}\BibitemShut {NoStop}%
\bibitem [{\citenamefont {Joseph}, \citenamefont {Beavers},\ and\ \citenamefont
  {Fosdick}(1973)}]{joseph1973free}%
  \BibitemOpen
  \bibfield  {author} {\bibinfo {author} {\bibfnamefont {D.~D.}\ \bibnamefont
  {Joseph}}, \bibinfo {author} {\bibfnamefont {G.~S.}\ \bibnamefont {Beavers}},
  \ and\ \bibinfo {author} {\bibfnamefont {R.~L.}\ \bibnamefont {Fosdick}},\
  }\bibfield  {title} {\enquote {\bibinfo {title} {The free surface on a liquid
  between cylinders rotating at different speeds part ii},}\ }\href@noop {}
  {\bibfield  {journal} {\bibinfo  {journal} {Archive for Rational Mechanics
  and Analysis}\ }\textbf {\bibinfo {volume} {49}},\ \bibinfo {pages}
  {381--401} (\bibinfo {year} {1973})}\BibitemShut {NoStop}%
\bibitem [{\citenamefont {Chandra}\ \emph {et~al.}(2021)\citenamefont
  {Chandra}, \citenamefont {Ghosh}, \citenamefont {Saha},\ and\ \citenamefont
  {Kumar}}]{chandra2021contact}%
  \BibitemOpen
  \bibfield  {author} {\bibinfo {author} {\bibfnamefont {N.~K.}\ \bibnamefont
  {Chandra}}, \bibinfo {author} {\bibfnamefont {U.~U.}\ \bibnamefont {Ghosh}},
  \bibinfo {author} {\bibfnamefont {A.}~\bibnamefont {Saha}}, \ and\ \bibinfo
  {author} {\bibfnamefont {A.}~\bibnamefont {Kumar}},\ }\bibfield  {title}
  {\enquote {\bibinfo {title} {Contact line pinning and depinning can modulate
  the rod-climbing effect},}\ }\href@noop {} {\bibfield  {journal} {\bibinfo
  {journal} {Langmuir}\ }\textbf {\bibinfo {volume} {37}},\ \bibinfo {pages}
  {14785--14792} (\bibinfo {year} {2021})}\BibitemShut {NoStop}%
\bibitem [{\citenamefont {Berman}, \citenamefont {Bradford},\ and\
  \citenamefont {Lundgren}(1978)}]{berman1978two}%
  \BibitemOpen
  \bibfield  {author} {\bibinfo {author} {\bibfnamefont {A.}~\bibnamefont
  {Berman}}, \bibinfo {author} {\bibfnamefont {J.}~\bibnamefont {Bradford}}, \
  and\ \bibinfo {author} {\bibfnamefont {T.}~\bibnamefont {Lundgren}},\
  }\bibfield  {title} {\enquote {\bibinfo {title} {Two-fluid spin-up in a
  centrifuge},}\ }\href@noop {} {\bibfield  {journal} {\bibinfo  {journal}
  {Journal of Fluid Mechanics}\ }\textbf {\bibinfo {volume} {84}},\ \bibinfo
  {pages} {411--431} (\bibinfo {year} {1978})}\BibitemShut {NoStop}%
\bibitem [{\citenamefont {Fujimoto}\ and\ \citenamefont
  {Takeda}(2009)}]{fujimoto2009topology}%
  \BibitemOpen
  \bibfield  {author} {\bibinfo {author} {\bibfnamefont {S.}~\bibnamefont
  {Fujimoto}}\ and\ \bibinfo {author} {\bibfnamefont {Y.}~\bibnamefont
  {Takeda}},\ }\bibfield  {title} {\enquote {\bibinfo {title} {Topology changes
  of the interface between two immiscible liquid layers by a rotating lid},}\
  }\href@noop {} {\bibfield  {journal} {\bibinfo  {journal} {Physical Review
  E}\ }\textbf {\bibinfo {volume} {80}},\ \bibinfo {pages} {015304} (\bibinfo
  {year} {2009})}\BibitemShut {NoStop}%
\bibitem [{\citenamefont {Zheng}\ \emph {et~al.}(2021)\citenamefont {Zheng},
  \citenamefont {Wen}, \citenamefont {Sun},\ and\ \citenamefont
  {Bai}}]{zheng2021effects}%
  \BibitemOpen
  \bibfield  {author} {\bibinfo {author} {\bibfnamefont {W.}~\bibnamefont
  {Zheng}}, \bibinfo {author} {\bibfnamefont {B.}~\bibnamefont {Wen}}, \bibinfo
  {author} {\bibfnamefont {C.}~\bibnamefont {Sun}}, \ and\ \bibinfo {author}
  {\bibfnamefont {B.}~\bibnamefont {Bai}},\ }\bibfield  {title} {\enquote
  {\bibinfo {title} {Effects of surface wettability on contact line motion in
  liquid--liquid displacement},}\ }\href@noop {} {\bibfield  {journal}
  {\bibinfo  {journal} {Physics of Fluids}\ }\textbf {\bibinfo {volume} {33}},\
  \bibinfo {pages} {082101} (\bibinfo {year} {2021})}\BibitemShut {NoStop}%
\bibitem [{\citenamefont {Fetzer}, \citenamefont {Ramiasa},\ and\ \citenamefont
  {Ralston}(2009)}]{fetzer2009dynamics}%
  \BibitemOpen
  \bibfield  {author} {\bibinfo {author} {\bibfnamefont {R.}~\bibnamefont
  {Fetzer}}, \bibinfo {author} {\bibfnamefont {M.}~\bibnamefont {Ramiasa}}, \
  and\ \bibinfo {author} {\bibfnamefont {J.}~\bibnamefont {Ralston}},\
  }\bibfield  {title} {\enquote {\bibinfo {title} {Dynamics of liquid- liquid
  displacement},}\ }\href@noop {} {\bibfield  {journal} {\bibinfo  {journal}
  {Langmuir}\ }\textbf {\bibinfo {volume} {25}},\ \bibinfo {pages} {8069--8074}
  (\bibinfo {year} {2009})}\BibitemShut {NoStop}%
\bibitem [{\citenamefont {Ramiasa}\ \emph {et~al.}(2012)\citenamefont
  {Ramiasa}, \citenamefont {Ralston}, \citenamefont {Fetzer},\ and\
  \citenamefont {Sedev}}]{ramiasa2012nanoroughness}%
  \BibitemOpen
  \bibfield  {author} {\bibinfo {author} {\bibfnamefont {M.}~\bibnamefont
  {Ramiasa}}, \bibinfo {author} {\bibfnamefont {J.}~\bibnamefont {Ralston}},
  \bibinfo {author} {\bibfnamefont {R.}~\bibnamefont {Fetzer}}, \ and\ \bibinfo
  {author} {\bibfnamefont {R.}~\bibnamefont {Sedev}},\ }\bibfield  {title}
  {\enquote {\bibinfo {title} {Nanoroughness impact on liquid--liquid
  displacement},}\ }\href@noop {} {\bibfield  {journal} {\bibinfo  {journal}
  {The Journal of Physical Chemistry C}\ }\textbf {\bibinfo {volume} {116}},\
  \bibinfo {pages} {10934--10943} (\bibinfo {year} {2012})}\BibitemShut
  {NoStop}%
\bibitem [{\citenamefont {Zanini}\ \emph {et~al.}(2017)\citenamefont {Zanini},
  \citenamefont {Marschelke}, \citenamefont {Anachkov}, \citenamefont {Marini},
  \citenamefont {Synytska},\ and\ \citenamefont {Isa}}]{zanini2017universal}%
  \BibitemOpen
  \bibfield  {author} {\bibinfo {author} {\bibfnamefont {M.}~\bibnamefont
  {Zanini}}, \bibinfo {author} {\bibfnamefont {C.}~\bibnamefont {Marschelke}},
  \bibinfo {author} {\bibfnamefont {S.~E.}\ \bibnamefont {Anachkov}}, \bibinfo
  {author} {\bibfnamefont {E.}~\bibnamefont {Marini}}, \bibinfo {author}
  {\bibfnamefont {A.}~\bibnamefont {Synytska}}, \ and\ \bibinfo {author}
  {\bibfnamefont {L.}~\bibnamefont {Isa}},\ }\bibfield  {title} {\enquote
  {\bibinfo {title} {Universal emulsion stabilization from the arrested
  adsorption of rough particles at liquid-liquid interfaces},}\ }\href@noop {}
  {\bibfield  {journal} {\bibinfo  {journal} {Nature communications}\ }\textbf
  {\bibinfo {volume} {8}},\ \bibinfo {pages} {1--9} (\bibinfo {year}
  {2017})}\BibitemShut {NoStop}%
\bibitem [{\citenamefont {Hejazi}\ and\ \citenamefont
  {Nosonovsky}(2013)}]{hejazi2013contact}%
  \BibitemOpen
  \bibfield  {author} {\bibinfo {author} {\bibfnamefont {V.}~\bibnamefont
  {Hejazi}}\ and\ \bibinfo {author} {\bibfnamefont {M.}~\bibnamefont
  {Nosonovsky}},\ }\bibfield  {title} {\enquote {\bibinfo {title} {Contact
  angle hysteresis in multiphase systems},}\ }\href@noop {} {\bibfield
  {journal} {\bibinfo  {journal} {Colloid and Polymer Science}\ }\textbf
  {\bibinfo {volume} {291}},\ \bibinfo {pages} {329--338} (\bibinfo {year}
  {2013})}\BibitemShut {NoStop}%
\bibitem [{\citenamefont {Fermigier}\ and\ \citenamefont
  {Jenffer}(1991)}]{fermigier1991experimental}%
  \BibitemOpen
  \bibfield  {author} {\bibinfo {author} {\bibfnamefont {M.}~\bibnamefont
  {Fermigier}}\ and\ \bibinfo {author} {\bibfnamefont {P.}~\bibnamefont
  {Jenffer}},\ }\bibfield  {title} {\enquote {\bibinfo {title} {An experimental
  investigation of the dynamic contact angle in liquid-liquid systems},}\
  }\href@noop {} {\bibfield  {journal} {\bibinfo  {journal} {Journal of colloid
  and interface science}\ }\textbf {\bibinfo {volume} {146}},\ \bibinfo {pages}
  {226--241} (\bibinfo {year} {1991})}\BibitemShut {NoStop}%
\bibitem [{\citenamefont {Seveno}\ \emph {et~al.}(2011)\citenamefont {Seveno},
  \citenamefont {Blake}, \citenamefont {Goossens},\ and\ \citenamefont
  {De~Coninck}}]{seveno2011predicting}%
  \BibitemOpen
  \bibfield  {author} {\bibinfo {author} {\bibfnamefont {D.}~\bibnamefont
  {Seveno}}, \bibinfo {author} {\bibfnamefont {T.}~\bibnamefont {Blake}},
  \bibinfo {author} {\bibfnamefont {S.}~\bibnamefont {Goossens}}, \ and\
  \bibinfo {author} {\bibfnamefont {J.}~\bibnamefont {De~Coninck}},\ }\bibfield
   {title} {\enquote {\bibinfo {title} {Predicting the wetting dynamics of a
  two-liquid system},}\ }\href@noop {} {\bibfield  {journal} {\bibinfo
  {journal} {Langmuir}\ }\textbf {\bibinfo {volume} {27}},\ \bibinfo {pages}
  {14958--14967} (\bibinfo {year} {2011})}\BibitemShut {NoStop}%
\bibitem [{\citenamefont {Brackbill}, \citenamefont {Kothe},\ and\
  \citenamefont {Zemach}(1992)}]{brackbill1992continuum}%
  \BibitemOpen
  \bibfield  {author} {\bibinfo {author} {\bibfnamefont {J.~U.}\ \bibnamefont
  {Brackbill}}, \bibinfo {author} {\bibfnamefont {D.~B.}\ \bibnamefont
  {Kothe}}, \ and\ \bibinfo {author} {\bibfnamefont {C.}~\bibnamefont
  {Zemach}},\ }\bibfield  {title} {\enquote {\bibinfo {title} {A continuum
  method for modeling surface tension},}\ }\href@noop {} {\bibfield  {journal}
  {\bibinfo  {journal} {Journal of computational physics}\ }\textbf {\bibinfo
  {volume} {100}},\ \bibinfo {pages} {335--354} (\bibinfo {year}
  {1992})}\BibitemShut {NoStop}%
\bibitem [{\citenamefont {Joseph}, \citenamefont {Nguyen},\ and\ \citenamefont
  {Beavers}(1984)}]{joseph_nguyen_beavers_1984}%
  \BibitemOpen
  \bibfield  {author} {\bibinfo {author} {\bibfnamefont {D.~D.}\ \bibnamefont
  {Joseph}}, \bibinfo {author} {\bibfnamefont {K.}~\bibnamefont {Nguyen}}, \
  and\ \bibinfo {author} {\bibfnamefont {G.~S.}\ \bibnamefont {Beavers}},\
  }\bibfield  {title} {\enquote {\bibinfo {title} {Non-uniqueness and stability
  of the configuration of flow of immiscible fluids with different
  viscosities},}\ }\href {\doibase 10.1017/S0022112084000872} {\bibfield
  {journal} {\bibinfo  {journal} {Journal of Fluid Mechanics}\ }\textbf
  {\bibinfo {volume} {141}},\ \bibinfo {pages} {319–345} (\bibinfo {year}
  {1984})}\BibitemShut {NoStop}%
\bibitem [{\citenamefont {Joseph}\ \emph {et~al.}(1985)\citenamefont {Joseph},
  \citenamefont {Renardy}, \citenamefont {Renardy},\ and\ \citenamefont
  {Nguyen}}]{joseph_renardy_renardy_nguyen_1985}%
  \BibitemOpen
  \bibfield  {author} {\bibinfo {author} {\bibfnamefont {D.~D.}\ \bibnamefont
  {Joseph}}, \bibinfo {author} {\bibfnamefont {Y.}~\bibnamefont {Renardy}},
  \bibinfo {author} {\bibfnamefont {M.}~\bibnamefont {Renardy}}, \ and\
  \bibinfo {author} {\bibfnamefont {K.}~\bibnamefont {Nguyen}},\ }\bibfield
  {title} {\enquote {\bibinfo {title} {Stability of rigid motions and rollers
  in bicomponent flows of immiscible liquids},}\ }\href {\doibase
  10.1017/S0022112085001185} {\bibfield  {journal} {\bibinfo  {journal}
  {Journal of Fluid Mechanics}\ }\textbf {\bibinfo {volume} {153}},\ \bibinfo
  {pages} {151–165} (\bibinfo {year} {1985})}\BibitemShut {NoStop}%
\end{thebibliography}%

\end{document}